\documentclass[aps,prd,superscriptaddress,nofootinbib,amsmath,amsfonts,preprintnumbers,groupedaddress,showpacs,twocolumn, 10pt,english]{revtex4-2}
\usepackage{graphicx}
\usepackage{tabularx}
\usepackage{array}
\newcolumntype{C}[1]{>{\centering\arraybackslash}p{#1}}
\usepackage{amsmath}
\usepackage{amssymb}
\usepackage{babel}
\usepackage{wrapfig}
\usepackage{cancel}

\usepackage{relsize,exscale}
\makeatletter

\usepackage{array,multirow,graphicx}
\usepackage{soul}
\usepackage{dcolumn}
\usepackage{newlfont}
\usepackage{bm}
\usepackage[colorlinks,citecolor=blue,urlcolor=blue,linkcolor=blue]{hyperref}
\usepackage[figtopcap]{subfigure}
\usepackage{color}
\usepackage{verbatim}

\renewcommand{\d}{\delta}

\def\be{\begin{align}}
\def\ee{\end{align}}
\def\bea{\begin{eqnarray}}
\def\eea{\end{eqnarray}}
\def\bal{\begin{align}}
\def\eal{\end{align}}

\usepackage{scalerel}
\usepackage{tikz}
\usetikzlibrary{svg.path}
\definecolor{orcidlogocol}{HTML}{A6CE39}
\tikzset{
 orcidlogo/.pic={
 \fill[orcidlogocol] svg{M256,128c0,70.7-57.3,128-128,128C57.3,256,0,198.7,0,128C0,57.3,57.3,0,128,0C198.7,0,256,57.3,256,128z};
 \fill[white] svg{M86.3,186.2H70.9V79.1h15.4v48.4V186.2z}
 svg{M108.9,79.1h41.6c39.6,0,57,28.3,57,53.6c0,27.5-21.5,53.6-56.8,53.6h-41.8V79.1z M124.3,172.4h24.5c34.9,0,42.9-26.5,42.9-39.7c0-21.5-13.7-39.7-43.7-39.7h-23.7V172.4z}
 svg{M88.7,56.8c0,5.5-4.5,10.1-10.1,10.1c-5.6,0-10.1-4.6-10.1-10.1c0-5.6,4.5-10.1,10.1-10.1C84.2,46.7,88.7,51.3,88.7,56.8z};}}
\newcommand\orcid[1]{\href{https://orcid.org/#1}{\mbox{\scalerel*{
\begin{tikzpicture}[yscale=-1,transform shape]
\pic{orcidlogo};
\end{tikzpicture}
}{|}}}}
\graphicspath{{./}{Figs/}}
\begin{document}

\date{\today}

\title{Analytic semiclassical backreaction of a Schwarzschild black hole in a finite cavity: horizon shift, temperature renormalization, and canonical stability in the Hartle-Hawking State}
\author{G.G.L. Nashed$^{1}$}\email{nashed@bue.edu.eg}
\author{Alnadhief H. A. Alfedeel$^{2}$}\email{aaalnadhief@imamu.edu.sa}
\author{Tiberiu Harko$^{3,4}$}\email{tiberiu.harko@aira.astro.ro}
\affiliation{$^{1}$Centre for Theoretical Physics, The British University, P.O. Box 43, El Sherouk City, Cairo 11837, Egypt\\$^{2}$Department of Mathematics and Statistics, Imam Mohammad Ibn Saud Islamic University (IMSIU), Riyadh, 13318, Saudi Arabia\\$^{3}$Faculty of Physics, Babe\c s-Bolyai University, 1 Kog\u alniceanu Street, 400084 Cluj-Napoca, Romania\\$^{4}$Astronomical Observatory, 19 Cire\c silor Street, 400487 Cluj-Napoca, Romania}
\begin{abstract}
We construct an analytic model of static semiclassical backreaction for a Schwarzschild black hole in the Hartle--Hawking state enclosed within a finite spherical cavity. Using a minimal renormalized stress--energy tensor consistent with conservation, thermal asymptotics, and horizon regularity, we integrate the reduced semiclassical Einstein equations under Dirichlet boundary conditions at the cavity wall. This yields explicit expressions for the corrections to the mass function, redshift factor, horizon location, and surface gravity. We obtain a closed-form first-order correction to the Hawking temperature in terms of a dimensionless backreaction parameter and the cavity radius. The temperature shift decomposes into redshift renormalization, geometric horizon displacement, and a local energy-density contribution at the horizon. The perturbative expansion is controlled by a parameter of order $M_P^2/M^2$, ensuring validity for macroscopic black holes. The near-horizon geometry retains its universal Rindler$^{2}\times S^{2}$ structure, indicating that semiclassical effects renormalize rather than modify the geometric origin of Hawking radiation.

\end{abstract}

\maketitle

\section{Introduction}

Semiclassical gravity provides the leading framework for describing quantum effects of matter fields on classical spacetime geometry when curvature scales remain well below the Planck scale \cite{Birrell:1982ix,Parker:2009uva,Wald:1984rg,Frolov:1998wf,Wald:1995yp}.
In this approach the Einstein equations are sourced by the renormalized expectation value of the quantum stress-energy tensor, incorporating vacuum polarization while neglecting metric fluctuations of genuinely quantum gravitational origin \cite{Birrell:1982ix,Parker:2009uva}.
The geometric and thermodynamic foundations of black hole physics were established in the classical laws of black hole mechanics and the discovery of Hawking radiation \cite{Hawking:1974rv,Bardeen:1973gs,Hawking:1975vcx,Gibbons:1976ue}.
Subsequent numerical and analytic computations of the renormalized stress tensor in Schwarzschild spacetime firmly established the structure of semiclassical backreaction \cite{Anderson:1994hg,Anderson:1989vg,Candelas:1980zt,Howard:1984qp,Page:1982fm,Howard:1984ttx}.

The renormalized stress-energy tensor depends crucially on the quantum state \cite{Birrell:1982ix,Boulware:1974dm,Unruh:1976db,Hartle:1976tp}.
The Boulware vacuum is regular at infinity but singular at the horizon \cite{Boulware:1974dm}.
The Unruh state describes an evaporating black hole with outgoing flux \cite{Unruh:1976db,fabbri2005modeling}.
The Hartle-Hawking state represents thermal equilibrium and is regular on both past and future horizons \cite{Hartle:1976tp,Candelas:1980zt,Howard:1984qp}.
Its detailed structure has been computed numerically and modeled analytically in several studies \cite{Anderson:1994hg,frolov2011blackhole,Anderson:1989vg,Candelas:1980zt,Howard:1984qp,Page:1982fm,Howard:1984ttx}.

Infrared issues in asymptotically flat black hole thermodynamics were clarified by enclosing the black hole in a finite cavity \cite{York:1986it,York:1984wp}.
The quasilocal energy formalism provides a systematic boundary framework for defining thermodynamic quantities \cite{Brown:1992br}.
Static semiclassical backreaction within this finite-cavity context has been explored extensively \cite{York:1986it,Anderson:1994hg,Brown:1992br,York:1984wp}.
The conceptual consistency of Killing horizons and black hole mechanics rests on foundational works in gravitational thermodynamics \cite{Wald:1984rg,Frolov:1998wf,Wald:1995yp,Bardeen:1973gs,Hawking:1975vcx,Gibbons:1976ue}.

The canonical ensemble formulation of black hole thermodynamics in a finite cavity, originally developed by York~\cite{York:1986it,York:1984wp}, provides a consistent framework for discussing equilibrium and stability in asymptotically flat spacetimes. Semiclassical extensions of this framework, incorporating quantum stress--energy backreaction in the Hartle--Hawking state, have been investigated in various contexts, often relying on Page-type approximations~\cite{Page:1982fm} and numerical implementations of the renormalized stress tensor~\cite{Candelas:1980zt,Howard:1984qp,Anderson:1989vg,Anderson:1994hg}. In such treatments, thermodynamic quantities including energy, entropy, and specific heat can be computed within the canonical ensemble.

In the present work, rather than introducing a detailed numerical approximation for the renormalized stress tensor, we construct a minimal analytic Hartle--Hawking RSET (Renormalized Stress Energy Tensor) model, constrained only by conservation, thermal asymptotics, and horizon regularity, respectively. This allows the semiclassical Einstein equations to be integrated in closed form under Dirichlet boundary conditions at a finite cavity wall. As a result, we obtain explicit analytic expressions for the semiclassical corrections to the Hawking temperature and, consequently, to the canonical heat capacity at fixed cavity radius.

This analytic structure permits a direct evaluation of the semiclassical shift in the canonical stability threshold as a function of the cavity parameter
\begin{equation}
x_B = \frac{2GM}{r_B},
\end{equation}
and of the RSET parameters appearing in the minimal model. In contrast to previous treatments where stability properties are extracted numerically or implicitly through free-energy analysis, the present formulation provides a transparent analytic laboratory for isolating the geometric origin of quantum corrections to equilibrium thermodynamics in black hole spacetimes.

The Hawking temperature is geometrically determined by the surface gravity of a Killing horizon \cite{Hawking:1974rv,Bardeen:1973gs,Hawking:1975vcx,Gibbons:1976ue}.
Near any non-extremal horizon, the geometry reduces to Rindler space times a transverse sphere, ensuring universality of the thermal behavior \cite{Wald:1984rg,Frolov:1998wf,Wald:1995yp,Bardeen:1973gs}.
Numerical semiclassical corrections to the Hawking temperature were investigated in several detailed studies \cite{Anderson:1994hg,Anderson:1989vg,Candelas:1980zt,Page:1982fm,Howard:1984ttx}.
Higher-derivative stability issues in semiclassical gravity were addressed through order-reduction methods \cite{Simon:1990jn,Parker:1993dk}.

The present work constructs a controlled analytic model of static semiclassical backreaction for a Schwarzschild black hole in the Hartle-Hawking state enclosed within a finite cavity.
The model incorporates thermal asymptotics, conservation constraints, and near-horizon regularity \cite{Anderson:1994hg,Candelas:1980zt,Howard:1984qp,Page:1982fm,Howard:1984ttx}.
The resulting temperature correction decomposes into redshift renormalization, geometric horizon displacement, and a local slope correction determined by the horizon energy density \cite{Wald:1995yp,Bardeen:1973gs,Hawking:1975vcx}.
The perturbative control parameter aligns with Planck-suppressed scaling in semiclassical gravity \cite{Birrell:1982ix,Parker:2009uva,Wald:1995yp}. Unlike previous semiclassical treatments that rely primarily on numerical evaluations of the renormalized stress tensor, the present formulation yields closed-form analytic expressions for the horizon shift, temperature renormalization, and canonical stability correction. This analytic control makes transparent the geometric origin of semiclassical thermodynamic modifications and isolates the role of infrared boundary conditions.

The organization of this paper is as follows. In Sec.~\ref{II} we present the semiclassical framework for static, spherically symmetric spacetimes and derive the reduced Einstein equations governing the mass function and redshift factor. In Sec.~\ref{III} we review the regularity conditions defining the Hartle--Hawking state and formulate a parametrization of the stress tensor consistent with horizon regularity and conservation. In Sec.~\ref{IV} we introduce a minimal analytic model for the renormalized stress--energy tensor satisfying thermal asymptotics, conservation constraints, and finiteness at the horizon. In Sec.~\ref{V} we integrate the semiclassical field equations in a finite cavity with Dirichlet boundary conditions and obtain explicit expressions for the backreaction corrections. In Sec.~\ref{VI} we compute the semiclassical horizon shift, derive the first-order correction to the Hawking temperature, and determine the analytic shift of the canonical stability threshold. In Sec.~\ref{VII} we analyze the near-horizon geometry and demonstrate the persistence of the universal Rindler$^{2}\times S^{2}$ structure. In Sec.~\ref{VIII} we discuss the regime of validity of the semiclassical expansion. In Sec.~\ref{IX} we provide a physical interpretation of the temperature correction, and in Sec.~\ref{X} we summarize our results and outline possible extensions.
\section{Semiclassical Setup}\label{II}
\label{sec:semiclassical_setup}

We work within the framework of semiclassical gravity, in which the classical spacetime geometry is sourced by the renormalized expectation value of the quantum stress--energy tensor of matter fields. The semiclassical Einstein equation reads
\begin{equation}
G_{\mu\nu}[g] = 8\pi G \, \langle T_{\mu\nu} \rangle_{\text{ren}} ,
\label{eq:semiclassical_einstein}
\end{equation}
where $\langle T_{\mu\nu} \rangle_{\text{ren}}$ is computed in a specified quantum state on the background geometry and includes the appropriate renormalization counterterms.
Semiclassical gravity provides the leading-order description of quantum backreaction in the regime where curvature scales are small compared to the Planck scale
\cite{Birrell:1982ix,Wald:1995yp,Parker:2009uva}.

\paragraph{Metric Ansatz.} We restrict our attention to static, spherically symmetric geometries, which can be written in Schwarzschild-like coordinates as
\begin{equation}
ds^2 = -e^{2\psi(r)} f(r) \, dt^2
+ \frac{dr^2}{f(r)}
+ r^2 d\Omega^2 ,
\label{eq:metric_ansatz}
\end{equation}
where $d\Omega=d\theta ^2+sin^2\theta \d\phi^2$, where $(\theta,\phi)$ are the angular coordinates, and
\begin{equation}
f(r) = 1 - \frac{2Gm(r)}{r}.
\label{eq:f_definition}
\end{equation}
Here $m(r)$ is the generalized mass function and $\psi(r)$ encodes the redshift factor.
This parametrization is standard in analyses of static semiclassical backreaction
\cite{York:1986it,Anderson:1994hg,frolov2011blackhole}.

The classical Schwarzschild solution corresponds to
$m(r)=M$,
and
$\psi(r)=0$, respectively.

\paragraph{Energy-Stress Tensor Structure.} For a static quantum state compatible with spherical symmetry, the renormalized stress tensor takes the diagonal form
\begin{equation}
\langle T^{\mu}{}_{\nu} \rangle_{\text{ren}}
=
\mathrm{diag}\!\left(
-\rho(r), \,
p_r(r), \,
p_t(r), \,
p_t(r)
\right),
\label{eq:stress_tensor_structure}
\end{equation}
where $\rho$ is the energy density, $p_r$ the radial pressure, and $p_t$ the tangential pressure.

The energy-stress tensor satisfies the conservation law
\begin{equation}
\nabla_\mu \langle T^{\mu}{}_{\nu} \rangle_{\text{ren}} = 0,
\label{eq:conservation}
\end{equation}
provides one nontrivial differential relation among $\rho, p_r, p_t$ and plays an essential role in constructing consistent semiclassical sources.

\paragraph{Reduced Field Equations}

Substituting the metric ansatz \eqref{eq:metric_ansatz} into the semiclassical Einstein equation \eqref{eq:semiclassical_einstein}, one obtains the following reduced system of ordinary differential equations
\begin{align}
m'(r) &= 4\pi r^2 \rho(r),
\label{eq:m_equation} \\
\psi'(r) &=\frac{ 4\pi G r \left[ p_r(r) + \rho(r) \right]}{f(r)}.
\label{eq:psi_equation}
\end{align}

Equation ~\eqref{eq:m_equation} follows directly from the $tt$ component of the Einstein tensor and generalizes the Misner--Sharp mass relation to the semiclassical regime.
Equation \eqref{eq:psi_equation} arises from the difference between the $rr$ and $tt$ components and determines the redshift function $\psi(r)$. These equations form the standard starting point for the analyses of the static semiclassical black hole backreaction
\cite{York:1986it,Anderson:1994hg,frolov2011blackhole,Anderson:1989vg}. 

In this study, we are interested in the regime where quantum effects are small corrections to the classical Schwarzschild geometry. Introducing a dimensionless expansion parameter
\begin{equation}
\epsilon \sim \frac{G \hbar}{M^2},
\end{equation}
we expand
\begin{equation}
m(r) = M + \epsilon \, \delta m(r),
\qquad
\psi(r) = \epsilon \, \delta \psi(r),
\end{equation}
and evaluate $\langle T_{\mu\nu} \rangle_{\text{ren}}$ on the classical background to leading order.
This order-reduction procedure avoids spurious higher-derivative instabilities associated with curvature-squared counterterms
\cite{Simon:1990jn, Parker:1993dk}.  The resulting system provides a controlled semiclassical description of static backreaction in the large black hole regime $M \gg M_{\mathrm{Pl}}$.

\section{Hartle--Hawking Regularity}\label{III}
\label{sec:HH_regularity}

\paragraph{Quantum States on Schwarzschild geometry.}
In curved spacetime, the renormalized stress tensor depends crucially on the choice of quantum state
\cite{Birrell:1982ix, Wald:1995yp}.
For Schwarzschild geometry, three canonical states are typically considered:

\begin{itemize}
\item The \textbf{Boulware state}, which is vacuum at infinity but singular at the horizon \cite{Boulware:1974dm}.
\item The \textbf{Unruh state}, describing an evaporating black hole with outgoing Hawking flux \cite{Unruh:1976db}.
\item The \textbf{Hartle--Hawking (HH) state}, representing a black hole in thermal equilibrium with a heat bath at the Hawking temperature \cite{Hartle:1976tp}.
\end{itemize}

In this work we focus on the Hartle--Hawking state, which is static, time-symmetric, and regular on both the future and past horizons.

paragraph{{Absence of Flux.} In a static orthonormal frame $(\hat t,\hat r,\hat\theta,\hat\phi)$ adapted to the metric
\eqref{eq:metric_ansatz}, equilibrium requires that there is no net radial energy flux.
Thus
\begin{equation}
T^{\hat t}{}_{\hat r} = 0.
\label{eq:no_flux}
\end{equation}

This condition distinguishes the HH state from the Unruh state, where
$T^{\hat t}{}_{\hat r} \neq 0$ due to Hawking radiation flux at infinity
\cite{Unruh:1976db,fabbri2005modeling}. In Schwarzschild-like coordinates, staticity and spherical symmetry imply that the stress tensor is diagonal
\begin{equation}
\langle T^{\mu}{}_{\nu} \rangle_{\text{ren}}
=
\mathrm{diag}(-\rho,\, p_r,\, p_t,\, p_t).
\end{equation}

\paragraph{Regularity at the Horizon.} Regularity of the HH state means that the stress tensor is finite in a freely falling frame at the horizon.
To further investigate this condition, one must use coordinates regular across the horizon, such as the ingoing Eddington--Finkelstein coordinates
\begin{equation}
v = t + r_*,
\qquad
dr_* = \frac{dr}{f(r)}.
\end{equation}

In these coordinates the metric becomes
\begin{equation}
ds^2 = -e^{2\psi} f \, dv^2 + 2 e^{\psi} dv\,dr + r^2 d\Omega^2,
\end{equation}
which is manifestly regular at $r=r_h$ provided $f(r_h)=0$ and $\psi(r_h)$ is finite.

The stress tensor components must remain finite in this coordinate system.
A necessary and sufficient condition for finiteness at the horizon is that
\cite{Candelas:1980zt,Howard:1984qp}
\begin{equation}
T^t{}_t(r_h) = T^r{}_r(r_h).
\end{equation}

Using the identification
\begin{equation}
\rho = -T^t{}_t,
\qquad
p_r = T^r{}_r,
\label{eq:horizon_condition}
\end{equation}
this condition becomes
\begin{equation}
\rho(r_h) + p_r(r_h) = 0.
\end{equation}
More exactly, near the horizon where
\begin{equation}
f(r) \sim f'(r_h)(r-r_h),
\end{equation}
the regularity requirement strengthens to
\begin{equation}
\rho + p_r = \mathcal{O}(f),
\label{eq:regularity_condition}
\end{equation}
so that the potentially divergent term in the redshift equation $\psi'(r) = 4\pi G \frac{r(\rho+p_r)}{f(r)}$
remains finite. This cancellation mechanism is a standard feature of the HH stress tensor
\cite{Candelas:1980zt, Anderson:1994hg}.

To implement condition \eqref{eq:regularity_condition} explicitly,
we decompose the radial pressure as
\begin{equation}
p_r(r) = -\rho(r) + f_0(r)\,\mathcal{K}(r), 
\label{eq:pr_decomposition}
\end{equation}
where 
\begin{equation}
  f_0(r) = 1 - \frac{2GM}{r},   
\end{equation}
is the background Schwarzschild lapse function, and $\mathcal{K}(r)$ is a finite function at the horizon. With this parametrization
\begin{equation}
\rho + p_r = f_0(r)\,\mathcal{K}(r),
\end{equation}
and hence the critical $1/f$ term in $\psi'(r)$ cancels identically, giving
\begin{equation}
\psi'(r) = 4\pi G\, r\, \mathcal{K}(r),
\end{equation}
which is manifestly finite at $r=r_h$.

Equation \eqref{eq:pr_decomposition} therefore provides a convenient and
state-compatible parametrization of the HH stress tensor consistent with
staticity, conservation, and horizon regularity.

\section{Analytical Model for the Hartle--Hawking 
Stress Tensor}\label{IV}
\label{sec:toy_RSET}

\paragraph{Dimensionless Variable and General Structure.} It is convenient to introduce the dimensionless compactified radial coordinate
\begin{equation}
x = \frac{2GM}{r},
\end{equation}
which maps the exterior region $r \in [2GM,\infty)$ to $x \in (0,1]$.
This variable is commonly used in analytic and numerical studies of
renormalized stress tensors in Schwarzschild spacetime
\cite{Page:1982fm, Anderson:1994hg}. In the Hartle--Hawking state, the renormalized stress tensor satisfies the conditions:

\begin{itemize}
\item Thermal behavior at spatial infinity,
\item Regularity at the horizon,
\item Conservation $\nabla_\mu T^{\mu}{}_{\nu}=0$,
\item Staticity and spherical symmetry.
\end{itemize}

Exact expressions for $\langle T_{\mu\nu}\rangle_{\text{ren}}$ are known only numerically
\cite{Candelas:1980zt, Anderson:1994hg}.
However, their qualitative structure strongly constrains admissible analytic approximations.

\paragraph{Thermal Asymptotics.} At spatial infinity, the Hartle--Hawking state approaches a thermal bath at the Hawking temperature
\cite{Hartle:1976tp}. Thus,
\begin{equation}
\rho(r \to \infty) = \rho_\infty,
\qquad
p_r = p_t = \frac{1}{3}\rho_\infty,
\end{equation}
where $\rho_\infty \propto T_H^4$ is the Stefan--Boltzmann energy density \cite{Landau:1980mil,Rybicki:2004hfl,Weinberg:2008zzc}.

To incorporate this behavior, we adopt the ultra-minimal polynomial ansatz
\begin{equation}
{
\rho(x) = \rho_\infty (1 + a x),
}
\label{eq:rho_ansatz}
\end{equation}
where $a$ is a dimensionless parameter encoding near-horizon deviations from pure thermal radiation.
Such low-order polynomial approximations have been successfully employed in analytic fits to numerical stress tensors
\cite{Page:1982fm,frolov2011blackhole}. The present minimal polynomial ansatz is not intended as a quantitative fit to numerical RSET data, but rather as a structurally consistent analytic model capturing the correct asymptotic, conservation, and regularity properties.

\subsection{Implementation of Hartle--Hawking Regularity}

As discussed in Section~\ref{sec:HH_regularity},
regularity at the horizon requires that Eq.~\eqref{eq:pr_decomposition} is satisfied.
We therefore parametrize the deviation from perfect vacuum-like behavior via
\begin{equation}
{
\mathcal{K}(x) = \frac{4}{3}\rho_\infty (1 + k x), 
}
\label{eq:K_ansatz}
\end{equation}
where $k$ is another dimensionless parameter.
The radial pressure is then constructed as
\begin{equation}
{
p_r(x) = -\rho(x) + (1-x)\mathcal{K}(x).
}
\label{eq:pr_ansatz}
\end{equation}

This guarantees that
\begin{equation}
\rho + p_r = (1-x)\mathcal{K}(x),
\end{equation}
which vanishes linearly at the horizon $x \to 1$, and ensures finiteness of $\psi'(r)$.

\paragraph{Conservation equation and determination of $p_t$.} The stress tensor must satisfy the conservation equation
\begin{equation}
\nabla_\mu T^{\mu}{}_{r} = 0.
\end{equation}

For the metric \eqref{eq:metric_ansatz}, this equation reduces to \cite{Anderson:1994hg}
\begin{equation}
\frac{dp_r}{dr}
+ \frac{f_0'}{2f_0}(\rho+p_r)
+ \frac{2}{r}(p_r - p_t)
= 0.
\end{equation}

Rewriting the above equation in terms of $x=2GM/r$, and using the mathematical relationships
\[
\frac{d}{dr} = -\frac{x^2}{2GM}\frac{d}{dx},
\qquad
f_0 = 1-x,
\]
one obtains after straightforward algebra the expression
\begin{equation}
{
p_t(x) =
p_r(x)
-
\frac{x}{2}\frac{dp_r}{dx}
+
\frac{x}{4}\mathcal{K}(x).
}
\label{eq:pt_solution}
\end{equation}

Thus, the tangential pressure is not an independent function;
it is uniquely determined by $\rho$ and $p_r$ through conservation.

This construction has the following properties:
\begin{itemize}
\item \textbf{Thermal limit at infinity:}
As $x \to 0$, one finds
\[
\rho \to \rho_\infty,
\qquad
p_r \to \frac{1}{3}\rho_\infty,
\qquad
p_t \to \frac{1}{3}\rho_\infty.
\]

\item \textbf{Hartle--Hawking regularity:}
$\rho + p_r \propto (1-x)$, guaranteeing horizon finiteness.

\item \textbf{Stress tensor conservation:}
Equation \eqref{eq:pt_solution} enforces $\nabla_\mu T^{\mu}{}_{\nu}=0$.

\item \textbf{Finite horizon behavior:}
All components remain finite as $x \to 1$.
\end{itemize}

The model therefore captures the essential structural properties of the exact Hartle--Hawking stress tensor, while remaining analytically tractable. It provides a controlled analytic laboratory for studying semiclassical backreaction.

\section{Backreaction in a Finite Cavity}\label{V}

To consistently incorporate semiclassical backreaction, {\it we consider the black hole enclosed in a finite spherical cavity of radius} $r=r_B$. The cavity provides a well-defined thermodynamic ensemble and avoids infrared divergences associated with asymptotically flat spacetimes \cite{York:1986it,Brown:1992br}.

We impose Dirichlet boundary conditions at the cavity wall \cite{York:1986it,Milton:2001yy}, fixing the induced metric there. In particular, we require that the first-order corrections to the mass function and redshift factor vanish at $r=r_B$
\begin{equation}
\delta m(r_B)=0,
\qquad
\delta\psi(r_B)=0.
\end{equation}
These conditions ensure that the total ADM mass and the proper normalization of the timelike Killing vector are fixed at the boundary.

\paragraph{Semiclassical Field Equations.} In semiclassical gravity, the Einstein equations take the form
\begin{equation}
G_{\mu\nu} = 8\pi G \langle T_{\mu\nu} \rangle ,
\end{equation}
where $\langle T_{\mu\nu} \rangle$ is the renormalized stress-energy tensor of quantum matter fields \cite{Birrell:1982ix,Parker:2009uva}.

For the  static, spherically symmetric metric given by Eq.~\eqref{eq:metric_ansatz},
we expand $m$ and $\psi$ around a classical background solution, thus obtaining
\begin{equation}
m(r)=M+\delta m(r),
\qquad
\psi(r)=\delta\psi(r),
\end{equation}
where $\delta m$ and $\delta\psi$ are first order in $\hbar$ \footnote{Hereafter we absorb the bookkeeping parameter, $\epsilon$,  into $\delta m$ and $\delta\psi$ and work to first order.}. To leading order in the semiclassical expansion, the $tt$ and $rr$ components of the Einstein equations reduce to \cite{York:1984wp,Anderson:1994hg}
\begin{eqnarray}
\delta m'(r) &= &4\pi r^2 \rho(r), 
\delta\psi'(r)= 4\pi G r \mathcal{K}(r), \nonumber\\
\mathcal{K}(r) &=& \langle T^{r}{}_{r} \rangle - \langle T^{t}{}_{t} \rangle, 
\end{eqnarray}
where $\rho(r) = - \langle T^{t}{}_{t} \rangle$. Note that
here $\rho(r)$ represents the local energy density of the quantum fields, while $\mathcal{K}(r)$ measures the anisotropy between radial pressure and energy density.

\paragraph{Integration of the backreaction equations.} Assuming that far from the horizon the quantum stress tensor approaches a constant asymptotic value
\begin{equation}
\rho(r) \simeq \rho_\infty
\left(1 + \frac{aGM}{r} \right),
\;\;
\mathcal{K}(r) \simeq
\frac{4}{3}\rho_\infty
\left(1 + \frac{kGM}{r} \right),
\end{equation}
where $a$ and $k$ are dimensionless constants determined by the quantum state and field content, we integrate the differential equations subject to the boundary conditions at $r=r_B$. 

Integrating $\delta m'(r)$ from $r$ to $r_B$ gives
\begin{equation}
\delta m(r)
= -4\pi\rho_\infty
\left[
\frac{r_B^3-r^3}{3}
+ aGM(r_B^2-r^2)
\right].
\end{equation}
The negative sign arises from enforcing $\delta m(r_B)=0$, ensuring that the total mass measured at the cavity boundary remains fixed.

Similarly, integrating the equation for $\delta\psi(r)$ yields
\begin{equation}
\delta\psi(r)
=
-\frac{16\pi G}{3}\rho_\infty
\left[
\frac{r_B^2-r^2}{2}
+2kGM(r_B-r)
\right].
\end{equation}
Again, the integration constant is chosen such that $\delta\psi(r_B)=0$, preserving the normalization of the Killing vector at the cavity wall.

\paragraph{Physical interpretation.} The above expressions explicitly demonstrate how quantum vacuum polarization inside the cavity, between the horizon and the boundary, modifies the effective mass function and the redshift factor. The corrections grow with the cavity size $r_B$, reflecting the cumulative contribution of the vacuum energy. The $r^3$ and $r^2$ terms correspond to a homogeneous energy density component, while the linear-in-$M$ terms encode curvature-induced corrections near the black hole.

Such finite-cavity constructions are standard in black hole thermodynamics and semiclassical backreaction analyses \cite{York:1986it,Brown:1992br,Anderson:1994hg}, providing a controlled framework for studying quantum corrections to black hole geometry.

\section{ Horizon Shift, Temperature Correction, 
and Canonical Stability}\label{VI}

We now analyze the semiclassical backreaction effects on the horizon position,
the Hawking temperature, and the canonical stability of a Schwarzschild black hole
enclosed in a finite cavity of radius $r_B$.

\subsection{Horizon position and Hawking temperature}

\paragraph{ Horizon condition and perturbative shift.} For the static, spherically symmetric metric of Eq.~\eqref{eq:metric_ansatz},
the event horizon is located at the largest root of
\begin{equation}
f(r_h)=0
\quad \Longrightarrow \quad
r_h = 2G m(r_h),
\end{equation}
a relation that remains valid in semiclassical gravity
\cite{Wald:1984rg,Frolov:1998wf}.

Expanding around the classical Schwarzschild solution
\begin{equation}
m(r)=M+\delta m(r),
\qquad
r_h = 2GM + \delta r_h,
\end{equation}
with $\delta m,\delta r_h=\mathcal{O}(\hbar)$, the horizon condition gives,
to first order
\begin{equation}
\delta r_h = 2G\,\delta m(2GM).
\end{equation}
This standard perturbative relation links the geometric displacement of the
horizon directly to the mass correction evaluated at the classical radius
\cite{York:1986it,Anderson:1994hg}.

It is convenient to introduce the dimensionless backreaction parameter
\begin{equation}\label{ms}
\eta = 32\pi G^2 M^2 \rho_\infty ,
\end{equation}
which measures the strength of quantum corrections relative to the classical curvature scale.
For a Hartle-Hawking state with $\rho_\infty \sim \hbar/M^4$, one finds
\begin{equation}
\eta \sim \frac{G\hbar}{M^2} \sim \frac{M_P^2}{M^2},
\end{equation}
so that in the semiclassical regime $M \gg M_P$ we have $\eta \ll 1$. We also define the cavity parameter
\begin{equation}
x_B = \frac{2GM}{r_B}.
\end{equation}

Using the finite-cavity solution for $\delta m(r)$, the fractional
horizon shift becomes
\begin{equation}
\frac{\delta r_h}{2GM}
=
-\eta
\left[
\frac{1}{3}\left(\frac{1}{x_B^3}-1\right)
+
\frac{a}{2}\left(\frac{1}{x_B^2}-1\right)
\right].
\label{hor_shift}
\end{equation}
The displacement grows as $x_B \to 0$ (large cavity),
reflecting the cumulative contribution of vacuum polarization energy
between the horizon and the boundary.

\paragraph{ Temperature correction.}
The Hawking temperature follows from the surface gravity
\begin{equation}
T_H = \frac{\kappa}{2\pi}
= \frac{1}{4\pi} e^{\psi_h} f'(r_h).
\end{equation}
Using
\begin{equation}
f'(r)
=
\frac{2G m(r)}{r^2}
-
\frac{2G m'(r)}{r},
\qquad
m'(r)=4\pi r^2 \rho(r),
\end{equation}
we obtain
\begin{equation}
T_H
=
\frac{e^{\psi_h}}{4\pi r_h}
\left(1 - 8\pi G r_h^2 \rho_h \right),
\end{equation}
where $\rho_h=\rho(r_h)$.
The temperature therefore receives both geometric and local
stress-energy contributions.

Expanding to first order relative to the classical value
$T_0=(8\pi GM)^{-1}$ yields
\begin{equation}
\frac{\delta T_H}{T_0}
=
\delta\psi(2GM)
-
\frac{\delta r_h}{2GM}
-
\eta(1+a).
\end{equation}
Substituting the explicit finite-cavity expressions gives
\begin{eqnarray}
\frac{\delta T_H}{T_0}
&=&
-\frac{\eta}{3}
\left[
\left(\frac{1}{x_B^2}-1\right)
+
2k\left(\frac{1}{x_B}-1\right)
\right]\nonumber\\
&&+
\eta
\left[
\frac{1}{3}\left(\frac{1}{x_B^3}-1\right)
+
\frac{a}{2}\left(\frac{1}{x_B^2}-1\right)
\right]
-
\eta(1+a).\nonumber\\
\label{temp_shift}
\end{eqnarray}

Thus the temperature shift receives three distinct contributions:
(i) redshift modification,
(ii) geometric horizon displacement,
and (iii) the explicit local energy-density term.
The dependence on $x_B$ demonstrates that semiclassical thermodynamics
is sensitive to the infrared regulator provided by the cavity.

\subsection{ Canonical stability in a cavity}

In the canonical ensemble at fixed $r_B$, stability is determined by
the wall temperature
\begin{equation}
T_B = \frac{T_H}{\sqrt{-g_{tt}(r_B)}},
\end{equation}
and the heat capacity
\begin{equation}
C_{r_B} =
\left( \frac{\partial M}{\partial T_B} \right)_{r_B}.
\end{equation}

Classically, the instability threshold is determined by
\begin{equation}
\left( \frac{\partial T_B}{\partial M} \right)_{r_B} = 0,
\end{equation}
which gives \cite{York:1986it}
\begin{equation}
x_B^{(0)} = \frac{2}{3}.
\end{equation}
Including semiclassical corrections we have
\begin{equation}
T_H(M)=T_0(M)\left[1+\eta F(x_B)\right],
\end{equation}
where $F(x_B)$ is defined by Eq.~\eqref{temp_shift}.
Expanding perturbatively we obtain
\begin{equation}
C_{r_B}
=
C_{r_B}^{(0)}
\left[
1
-
\eta\,\Delta(x_B;a,k)
\right]
+
\mathcal{O}(\eta^2),
\end{equation}
so that the canonical stability boundary shifts to
\begin{equation}
x_B^{\mathrm{crit}}
=
\frac{2}{3}
+
\eta\,\delta x_B(a,k)
+
\mathcal{O}(\eta^2).
\end{equation}

Using $\delta\psi(r_B)=0$ and $\delta m(r_B)=0$ we have
$-g_{tt}(r_B)=e^{2\psi(r_B)}f(r_B)=f(r_B)=1-2GM/r_B=1-x_B$, and therefore
\begin{equation}
T_B=\frac{T_H}{\sqrt{1-x_B}}.
\end{equation}

With $T_H(M)=T_0(M)\,[1+\eta F(x_B)]$ and $T_0=(8\pi GM)^{-1}$, one may write at fixed $r_B$
\begin{equation}
T_B(x)=\frac{1}{4\pi r_B}\,\frac{1+\eta F(x)}{x\sqrt{1-x}},
\qquad x\equiv x_B=\frac{2GM}{r_B}.
\end{equation}

The canonical stability threshold is determined by $(\partial T_B/\partial M)_{r_B}=0$, or,
equivalently, $dT_B/dx=0$. Expanding about the classical critical point $x_0=2/3$ with
$x_{\textrm crit}=x_0+\eta\,\delta x_B$ gives
\begin{equation}
\delta x_B=-\frac{A(x_0)F'(x_0)}{A''(x_0)},\qquad A(x)=\frac{1}{x\sqrt{1-x}}.
\end{equation}

Differentiating Eq.~(48) yields
\begin{equation}
F'(x)=\frac{2k x^2+x(2-3a)-3}{3x^4},
\end{equation}
and evaluating at $x_0=2/3$ results in the explicit shift
\begin{equation}
{\;\delta x_B(a,k)=\frac{a}{2}-\frac{2k}{9}+\frac{5}{12}\;}
\end{equation}
so that
\begin{equation}
x_B^{\textrm crit}=\frac{2}{3}+\eta\left(\frac{a}{2}-\frac{2k}{9}+\frac{5}{12}\right)+O(\eta^2).
\end{equation}

Therefore quantum backreaction renormalizes both the equilibrium
temperature and the canonical stability threshold.
The shift is controlled by the integrated properties of the
Hartle--Hawking stress tensor and vanishes smoothly in the
classical limit $\eta \to 0$.

\begin{figure}
\centering
\subfigure[~Normalized Horizon Shift vs Cavity Parameter]{\label{fig:1}\includegraphics[width=0.23\textwidth]{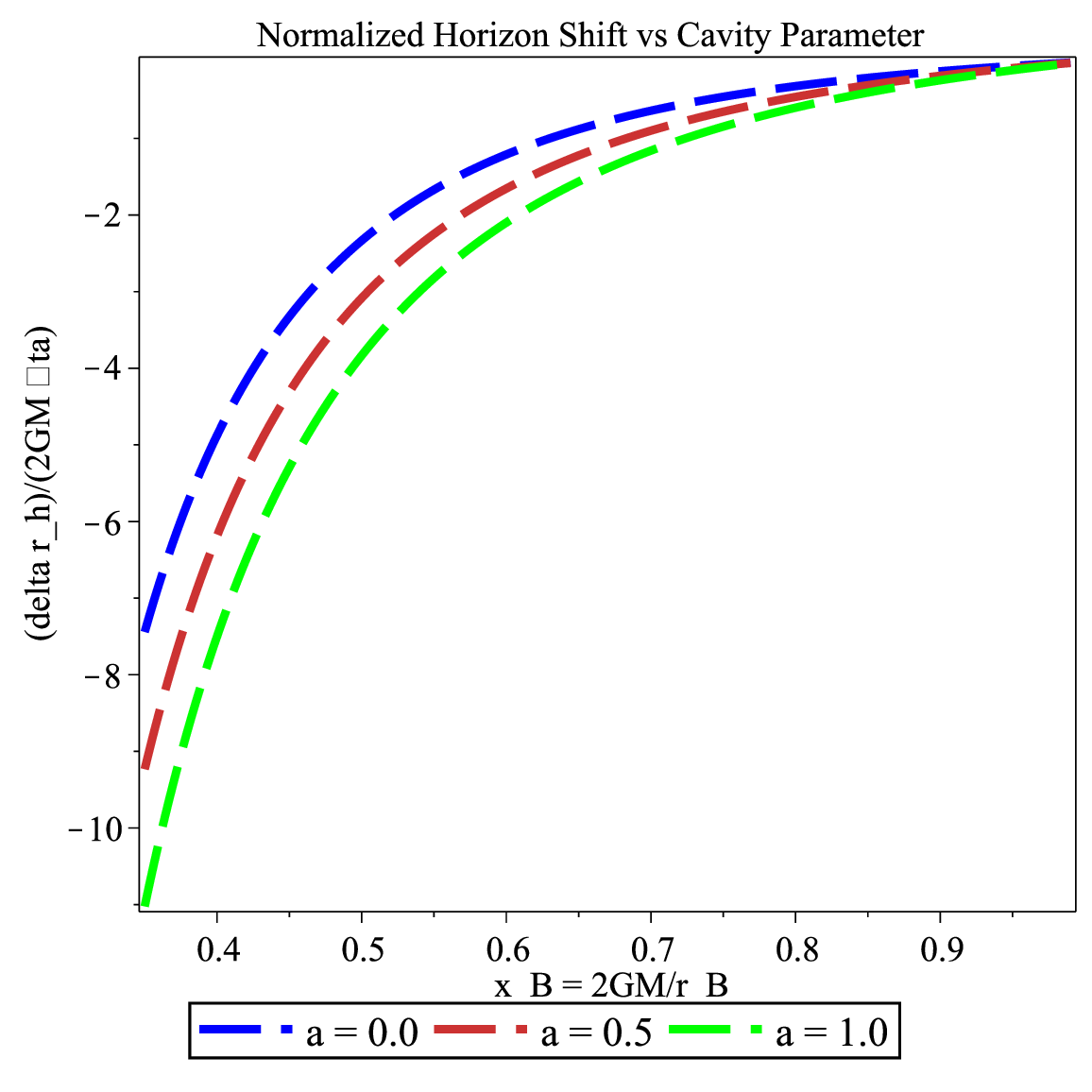}}
\subfigure[~Semiclassical Temperature Correction Function]{\label{fig:2}\includegraphics[width=0.23\textwidth]{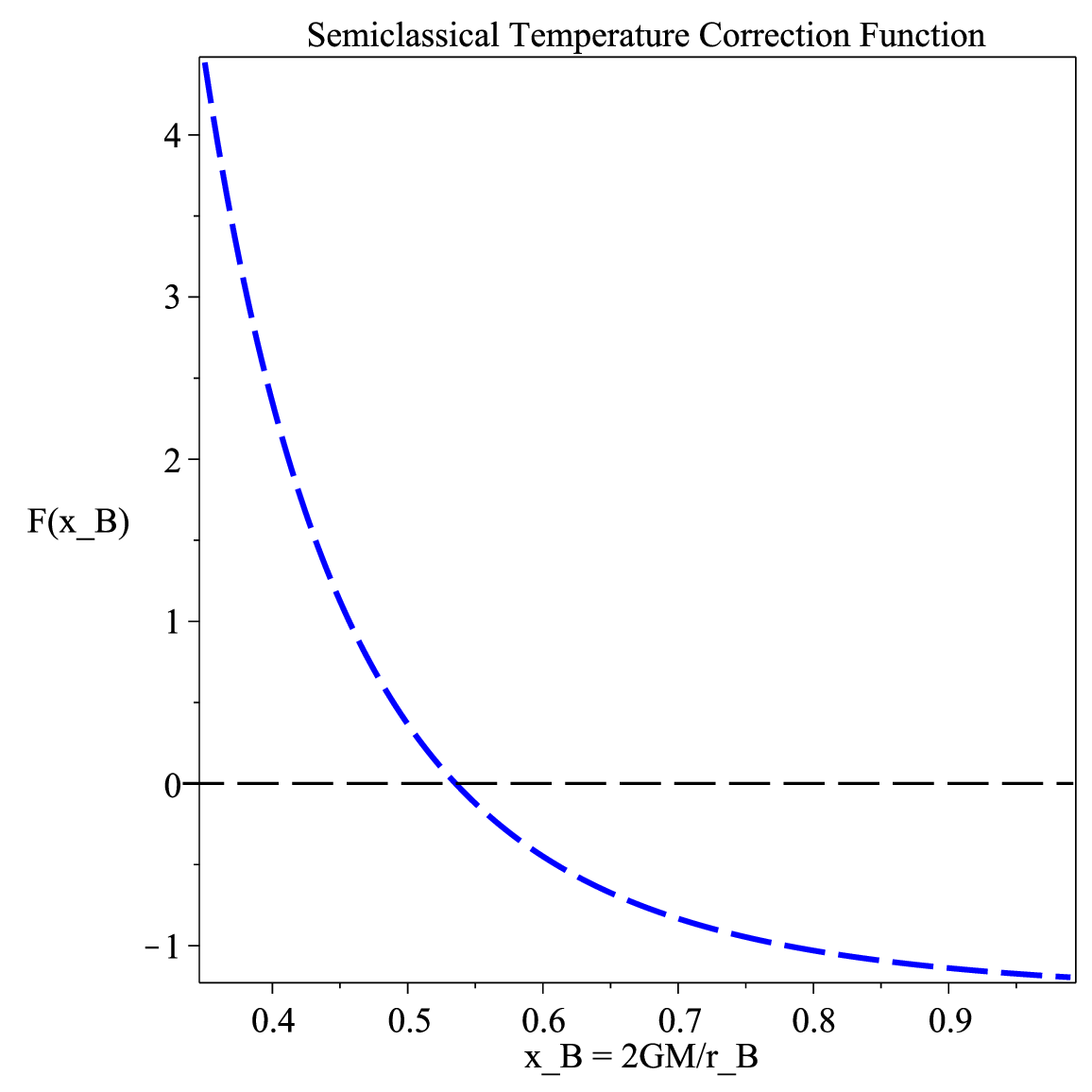}}
\subfigure[~Relative Semiclassical Correction to Heat Capacity]{\label{fig:3}\includegraphics[width=0.23\textwidth]{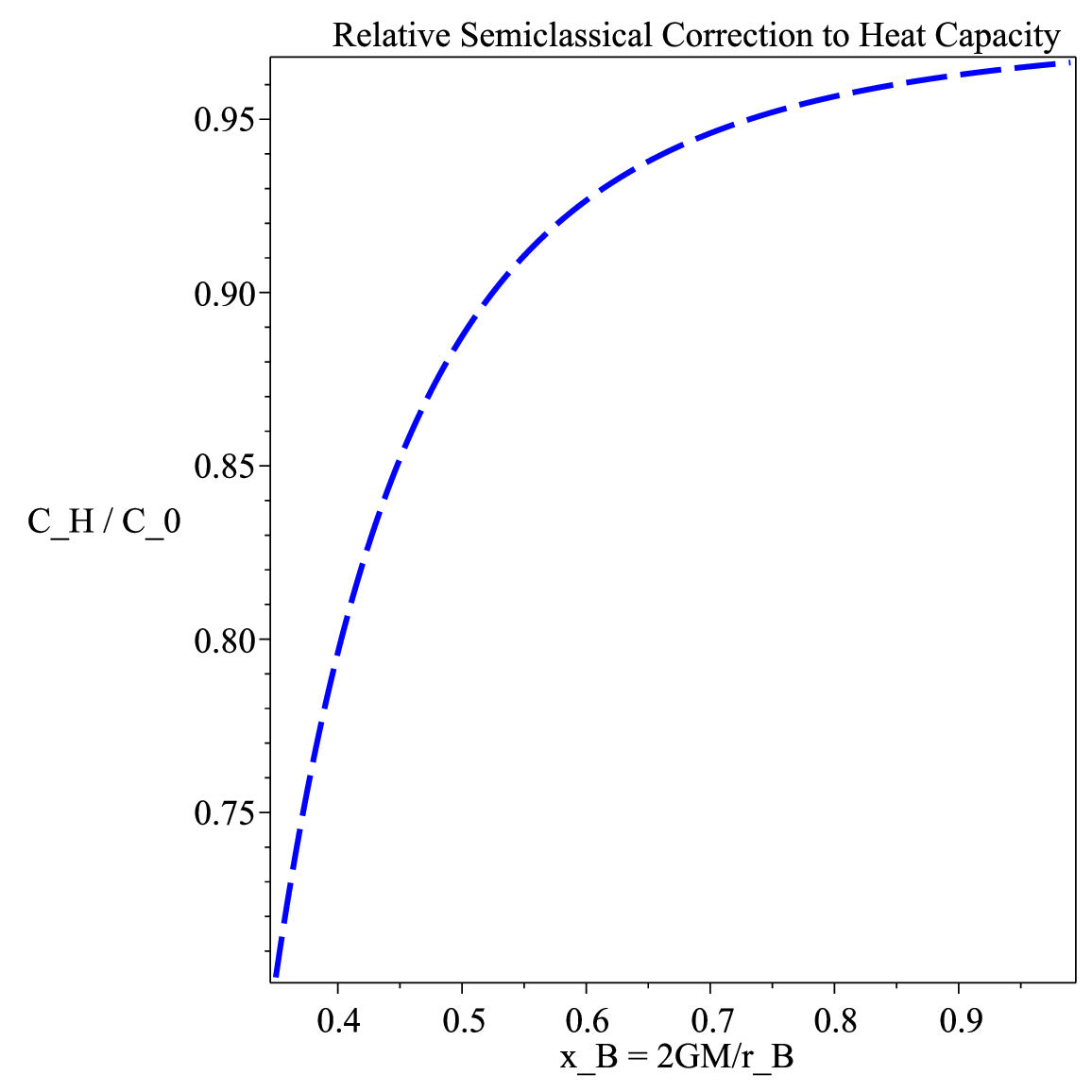}}
\subfigure[~Heat Capacity with Classical Stability Threshold]{\label{fig:4}\includegraphics[width=0.23\textwidth]{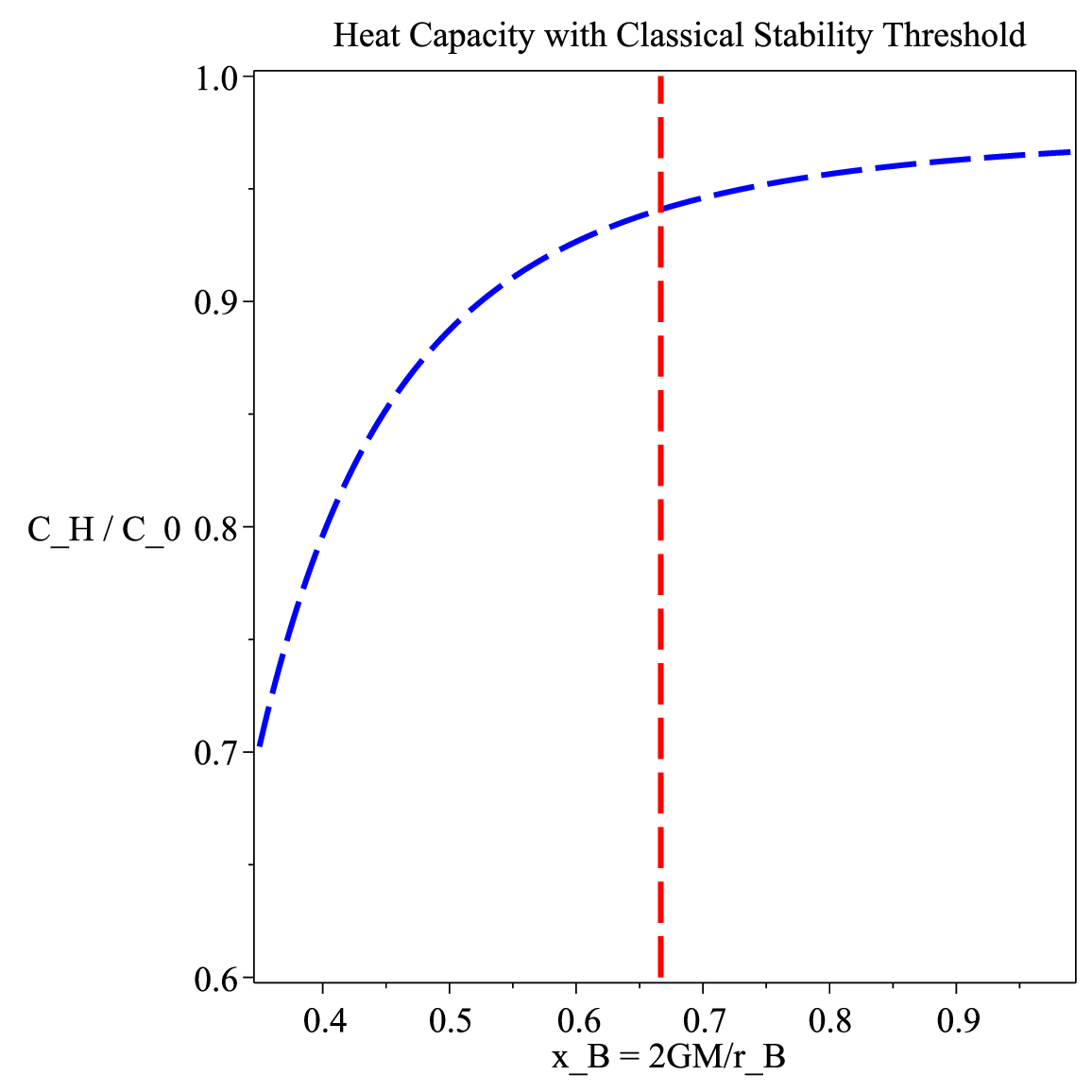}}
\caption{Canonical stability diagnostics for a Schwarzschild black hole in a finite cavity including semiclassical backreaction.
\subref{fig:1} Normalized horizon displacement as a function of the cavity parameter $x_B=2GM/r_B$ (shown for representative parameter choices in the stress-tensor model).
\subref{fig:2} Semiclassical temperature correction function $F(x_B)$ defined by $T_H(M)=T_0(M)\,[1+\eta F(x_B)]$.
\subref{fig:3} Relative semiclassical correction to the heat capacity at fixed cavity radius.
\subref{fig:4} Heat capacity including the classical stability threshold at $x_B=2/3$ (vertical dashed line), illustrating how semiclassical effects renormalize the response while preserving the qualitative stability structure.}
\label{fig:cavity}
\end{figure}

\paragraph{Physical analysis.} Figure~\ref{fig:cavity} illustrates the influence of semiclassical backreaction on the thermodynamic behavior of a Schwarzschild black hole enclosed in a finite cavity.\\
Panel \subref{fig:1} shows the normalized horizon shift as a function of the cavity parameter $x_B = 2GM/r_B$. The curve indicates that the horizon displacement remains small throughout the parameter range considered, confirming that semiclassical corrections act as perturbative modifications to the classical geometry. The monotonic behavior suggests that the cavity size controls the magnitude of backreaction effects, with stronger corrections emerging as the boundary approaches the gravitational radius.

Panel \subref{fig:2} presents the semiclassical correction function $F(x_B)$, which enters the modified Hawking temperature through $
T_H(M) = T_0(M)\,[1+\eta F(x_B)]$. The correction remains finite and smooth across the allowed range of $x_B$, indicating that the semiclassical modification does not introduce thermodynamic singularities. The functional form shows that the temperature shift becomes more pronounced near the classical stability threshold, reflecting enhanced sensitivity of the thermal state to quantum effects in this regime.

Panel \subref{fig:3} displays the relative semiclassical correction to the heat capacity at fixed cavity radius. Since the heat capacity determines canonical stability, this panel directly quantifies how quantum backreaction alters the response of the system to temperature fluctuations. The correction remains subleading as compared to the classical contribution, but it modifies the slope of the thermodynamic curve, effectively renormalizing the stability behavior.

Panel \subref{fig:4} shows the full heat capacity including the classical stability threshold at $x_B = 2/3$. The vertical dashed line marks the transition between thermodynamically unstable and stable branches. Importantly, semiclassical effects shift the quantitative values of the heat capacity while preserving the qualitative structure of the phase transition. This indicates that quantum corrections do not eliminate the canonical stabilization mechanism induced by the cavity, but rather adjust its precise location and strength.

Overall, Fig.~\ref{fig:cavity} demonstrates that semiclassical backreaction produces controlled, finite modifications to black hole thermodynamics. The corrections renormalize temperature and heat capacity without altering the fundamental stability structure of the canonical ensemble. This supports the interpretation that the cavity-regulated Schwarzschild system remains thermodynamically consistent under semiclassical effects, with quantum contributions acting as perturbative refinements rather than qualitative modifications of the classical picture.

\section{Near-Horizon Geometry}\label{VII}

Now we consider the static, spherically symmetric metric of Eq.~\eqref{eq:metric_ansatz} where the event horizon is located at the largest root $r=r_h$ of $f(r)$. 

\paragraph{Linear expansion around the horizon.} For a non-extremal black hole, the surface gravity is finite and
\begin{equation}
f'(r_h) \neq 0.
\end{equation}
In this case, $f(r)$ has a simple zero at $r_h$, and we may expand it linearly
\begin{equation}
f(r)
\approx
f'(r_h)(r-r_h)
+ \mathcal{O}\!\left((r-r_h)^2\right).
\end{equation}
Similarly, since $\psi(r)$ is regular at the horizon
\begin{equation}
\psi(r)
\approx
\psi_h
+
\mathcal{O}(r-r_h),
\qquad
\psi_h \equiv \psi(r_h).
\end{equation}

Such linear expansions are standard in analyses of black hole thermodynamics and near-horizon limits \cite{Wald:1984rg,Frolov:1998wf}.

To exhibit the regular geometry near the horizon, it is convenient to introduce a proper radial distance coordinate. From the metric, we obtain
\begin{equation}
dl^2 = \frac{dr^2}{f(r)}.
\end{equation}
Using the linear approximation
\begin{equation}
dl^2
\approx
\frac{dr^2}{f'(r_h)(r-r_h)}, 
\end{equation}
by defining a new coordinate $\rho$ through
\begin{equation} 
r - r_h
=
\frac{f'(r_h)}{4} \rho^2,
\end{equation}
and by taking into account that
\begin{equation}
dr
=
\frac{f'(r_h)}{2}\rho\, d\rho,
\end{equation}
we obtain
\begin{equation}
\frac{dr^2}{f(r)}
\approx
d\rho^2.
\end{equation}
Thus $\rho$ measures the proper distance from the horizon in the leading order.

\paragraph{Rindler form of the metric.}
Substituting the expansions into the full metric given by Eq.~\eqref{eq:metric_ansatz}, yields
\begin{eqnarray}
ds^2
&\approx&
- e^{2\psi_h} f'(r_h)(r-r_h) dt^2
+ d\rho^2
+ r_h^2 d\Omega^2 \nonumber\\
&=&
- \left( \frac{e^{\psi_h} f'(r_h)}{2} \right)^2
\rho^2 dt^2
+ d\rho^2
+ r_h^2 d\Omega^2.
\end{eqnarray}

Defining the surface gravity
\begin{equation}
\kappa
=
\frac{1}{2} e^{\psi_h} f'(r_h),
\end{equation}
which is the standard invariant definition for static spacetimes \cite{Wald:1995yp}, we obtain
\begin{equation}
ds^2
\approx
d\rho^2
-
\kappa^2 \rho^2 dt^2
+
r_h^2 d\Omega^2.
\end{equation}
This is precisely the direct product geometry $
\text{Rindler}_2 \times S^2$
where the two-dimensional $(\rho,t)$ sector is Rindler spacetime with acceleration $\kappa$, and the transverse sphere has fixed radius $r_h$.

The emergence of Rindler geometry near the horizon is universal for all non-extremal black holes \cite{Bardeen:1973gs}. It implies:

\begin{itemize}
\item The horizon is locally indistinguishable from an acceleration horizon.
\item Quantum field theory near the horizon reduces to QFT in Rindler space.
\item Regularity of the Euclidean section requires periodicity
\begin{equation}
t_E \sim t_E + \frac{2\pi}{\kappa},
\end{equation}
leading directly to the Hawking temperature \cite{Hawking:1975vcx,Gibbons:1976ue},
\begin{equation}
T_H = \frac{\kappa}{2\pi}.
\end{equation}

\end{itemize}

Thus, independently of global structure, the near-horizon region universally takes the form Rindler $\times S^2$, which underlies the robustness of black hole thermodynamics.

\section{Validity Regime}\label{VIII}

\paragraph{Perturbative Expansion Parameter.} Our analysis relies on a semiclassical expansion in which the spacetime
metric is written as
\begin{equation}
g_{\mu\nu}
=
g_{\mu\nu}^{(0)}
+
\hbar\, g_{\mu\nu}^{(1)}
+
\mathcal{O}(\hbar^2),
\end{equation}
where $g_{\mu\nu}^{(0)}$ is the classical Schwarzschild background and
$g_{\mu\nu}^{(1)}$ arises from the renormalized stress-energy tensor
$\langle T_{\mu\nu} \rangle$ of the quantum fields.
The semiclassical Einstein equations take the form
$
G_{\mu\nu}
=
8\pi G \langle T_{\mu\nu} \rangle$ ,
and they are valid when quantum fluctuations of the geometry remain small
as compared to the classical curvature scale
\cite{Birrell:1982ix,Parker:2009uva,Wald:1995yp}. 

In the finite-cavity setup, the strength of backreaction is governed by
the dimensionless parameter $\eta$, which is defined by Eq.~\eqref{ms},
where $\rho_\infty$ characterizes the asymptotic magnitude of the
renormalized energy density.

Perturbative control requires
\begin{equation}
{
\eta \ll 1,
\qquad
\eta(1+a) < 1.
}
\end{equation}

The first condition ensures that corrections to the horizon radius,
surface gravity, and redshift function remain small compared to their
classical values. The second condition guarantees that the corrected
temperature remains positive and that the horizon does not approach
extremality.

\paragraph{Scaling with the black hole mass}

Dimensional analysis clarifies the regime of validity of the present approach.
For quantum fields in a Hartle-Hawking-like state,
the typical magnitude of the renormalized stress tensor scales as
\cite{Howard:1984ttx,Anderson:1994hg}
\begin{equation}
\rho \sim \frac{\hbar}{M^4}.
\end{equation}
Substituting into $\eta$ gives
\begin{equation}
\eta
\sim
32\pi G^2 M^2 \frac{\hbar}{M^4}
\sim
\frac{\hbar G^2}{M^2}.
\end{equation}

Using $G\hbar = \ell_P^2$ and $M_P^2 = \hbar/G$,
this may be written as
\begin{equation}
{
\eta \sim \frac{\ell_P^2}{M^2}
=
\frac{M_P^2}{M^2}.
}
\end{equation}

Thus, for macroscopic black holes with
\begin{equation}
M \gg M_P,
\end{equation}
we have
\begin{equation}
\eta \ll 1,
\end{equation}
a condition indicating that semiclassical perturbation theory is well-controlled.

\paragraph{Physical interpretation.} The condition $\eta \ll 1$ is equivalent to requiring that
\begin{equation}
\frac{|\delta r_h|}{r_h} \ll 1,
\qquad
\frac{|\delta T_H|}{T_H} \ll 1,
\end{equation}
so that quantum backreaction produces only small fractional corrections.

As $M$ approaches the Planck scale
\begin{equation}
M \sim M_P,
\end{equation}
we find $\eta \sim 1$ and the perturbative expansion breaks down.
In this regime:

\begin{itemize}
\item Higher-loop corrections become important,
\item Metric fluctuations can no longer be neglected,
\item A full theory of quantum gravity is required.
\end{itemize}

This behavior is consistent with the standard expectation that
semiclassical gravity is reliable only when the curvature scale
$R \sim 1/M^2$ remains much smaller than the Planck scale
$1/\ell_P^2$ \cite{Wald:1995yp,Parker:2009uva}.

\paragraph{Cavity dependence.} The presence of the finite cavity does not alter the ultraviolet
validity condition $\eta \ll 1$, but it regulates infrared growth
of the integrated vacuum energy.
For fixed $M$, taking $r_B \to \infty$ increases the accumulated
backreaction and can eventually violate perturbative control,
reflecting the known infrared sensitivity of semiclassical
black hole thermodynamics \cite{York:1984wp,Brown:1992br}.

\paragraph{Summary of the validity regime.} 
The semiclassical solution derived above is self-consistent provided
\begin{equation}
{
M \gg M_P,
\qquad
\eta \ll 1,
\qquad
\eta(1+a) < 1.
}
\end{equation}

Within this regime, the backreacted geometry remains close to the
Schwarzschild one, the horizon is non-extremal, and the Rindler
near-horizon structure is preserved.

\section{Physical Interpretation}\label{IX}

The semiclassical correction to the Hawking temperature admits a
natural decomposition into three distinct contributions.

To first order in the backreaction parameter, the fractional
temperature shift may be written schematically as
\begin{equation}
\frac{\delta T_H}{T_0}
=
\underbrace{\delta\psi_h}_{\text{redshift}}
-
\underbrace{\frac{\delta r_h}{r_h}}_{\text{horizon shift}}
-
\underbrace{8\pi G r_h^2 \rho_h}_{\text{local slope correction}} .
\label{temperature-decomposition}
\end{equation}

Each term has a clear geometric and physical interpretation.
 \begin{figure}[t]
   \includegraphics[width=0.48\textwidth]{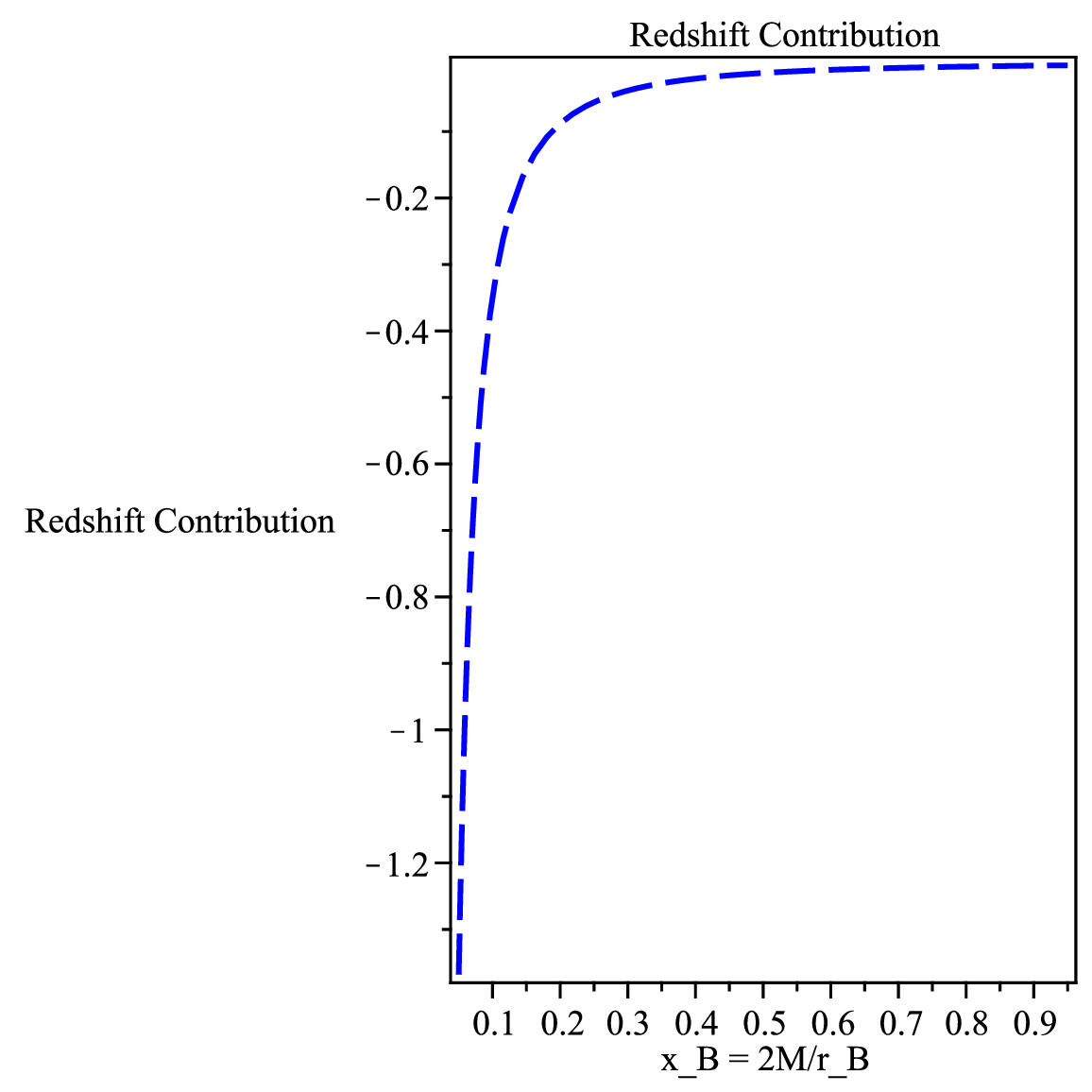}
    \caption{The redshift contribution $\delta\psi_h$ to $\delta T_H/T_0$ as a function of $x_B$ for representative values of $k$.}
  \label{fig:red}
\end{figure}

Figure~\ref{fig:red} displays the redshift contribution $\delta\psi_h$ to the fractional
temperature shift. This term arises from the modification of the lapse
function due to vacuum polarization outside the horizon.

Unlike the geometric horizon-shift term, the redshift correction is nonlocal:
it is obtained by integrating the anisotropy combination
$K = \langle T^r{}_r \rangle - \langle T^t{}_t \rangle$
between the horizon and the cavity boundary.

Its dependence on the cavity parameter $x_B$ therefore reflects the
accumulated semiclassical backreaction throughout the exterior region,
demonstrating that the normalization of the timelike Killing vector
is renormalized by quantum effects.

The factor $\delta\psi_h$ arises from the correction to the lapse
function evaluated at the horizon. Since the surface gravity is
\begin{equation}
\kappa = \frac{1}{2} e^{\psi_h} f'(r_h),
\end{equation}
any modification of the redshift factor $e^{\psi}$ directly rescales
the normalization of the timelike Killing vector.
This contribution depends on the anisotropy combination $\mathcal{K}$
and therefore on the parameter $k$ characterizing the large-$r$
structure of the renormalized stress-energy tensor (RSET).

Physically, this term reflects the fact that vacuum polarization
outside the horizon modifies the gravitational redshift between the
horizon and the cavity wall. It is a nonlocal effect, accumulated
through integration of the semiclassical Einstein equations
\cite{York:1986it,Anderson:1994hg}.

\paragraph{The  horizon shift.} The term $\delta r_h / r_h$ originates from the displacement of the
horizon due to the correction of the mass function. Since the
horizon satisfies the condition $
r_h = 2G m(r_h)$
a change in $m(r)$ shifts the location of the Killing horizon.

This contribution depends explicitly on the cavity parameter $x_B = \frac{2GM}{r_B}$
reflecting the integrated vacuum energy between the horizon and the
boundary. In other words, the horizon position responds to the total
enclosed energy, consistent with the quasilocal energy framework of
Brown and York \cite{Brown:1992br}.

Unlike the redshift term, this effect is geometric: it arises from
the condition defining the horizon itself and would be present even
if the lapse function were unmodified.
\paragraph{ Local Slope Correction.}
The final contribution,
\begin{equation}
-8\pi G r_h^2 \rho_h,
\end{equation}
arises directly from the derivative of $f(r)$ at the horizon
\begin{equation}
f'(r_h)
=
\frac{1}{r_h}
-
8\pi G r_h \rho_h.
\end{equation}

This term depends only on the local energy density
$\rho_h = -\langle T^t{}_t\rangle_{r_h}$ evaluated at the horizon.
It represents a purely local modification of the radial slope of the
metric function, and therefore of the surface gravity.

Physically, this reflects the fact that the Hawking temperature is
sensitive to the immediate near-horizon stress-energy content
\cite{Howard:1984ttx,Anderson:1994hg}.
It is independent of global boundary conditions and depends only on
the regularity of the quantum state at the horizon.

\paragraph{Structural nature of the decomposition.}
The separation (\ref{temperature-decomposition}) is structural and
does not rely on detailed numerical fits of the RSET. It follows
directly from:

\begin{enumerate}
\item The definition of surface gravity in static spacetimes
      \cite{Bardeen:1973gs,Wald:1995yp}.
\item The semiclassical relation $m'(r)=4\pi r^2\rho(r)$.
\item The perturbative expansion around a non-extremal horizon.
\end{enumerate}

Thus, independently of the specific field content or renormalization
scheme, any semiclassical correction to a static black hole temperature
must decompose into:

\begin{itemize}
\item A redshift (lapse) renormalization,
\item A geometric shift of the horizon,
\item A local stress-energy correction at the horizon.
\end{itemize}

This universality reflects the fact that the Hawking temperature is
determined entirely by near-horizon geometry. The Rindler structure
of the horizon persists under small backreaction, and quantum
corrections merely renormalize the surface gravity rather than
altering its geometric origin \cite{Hawking:1975vcx,Wald:1995yp}.

\section{Discussions and final remarks}\label{X}

In this work we have constructed a controlled analytic model of static semiclassical backreaction for a Schwarzschild black hole in the Hartle--Hawking state enclosed within a finite cavity. 

By combining thermal asymptotics, stress--tensor conservation, and near-horizon regularity, we obtained an analytically tractable representation of the renormalized stress--energy tensor that captures the essential structural properties of previously computed numerical results. 

The finite-cavity framework regulates infrared contributions and imposes well-defined boundary conditions, ensuring a consistent thermodynamic interpretation within a canonical ensemble.

The semiclassical field equations reduce to coupled differential equations for the mass function and redshift factor. Integrating these equations with Dirichlet boundary conditions at the cavity wall yields explicit expressions for the backreaction corrections. These corrections encode two physically distinct effects: a cumulative vacuum polarization contribution that grows with cavity size, and curvature-dependent contributions localized near the horizon. The resulting analytic form makes transparent how quantum vacuum energy perturbs the classical Schwarzschild geometry without requiring detailed numerical input.

A central result is the closed-form expression for the first-order shift of the Hawking temperature. The correction decomposes into three geometrically distinct components: (i) a redshift renormalization due to vacuum polarization outside the horizon, (ii) a geometric displacement of the Killing horizon arising from the corrected mass function, and (iii) a local slope correction determined by the energy density at the horizon. This decomposition follows directly from the definition of surface gravity and the perturbative expansion around a non-extremal horizon, and does not rely on detailed numerical fits to the stress tensor.

At leading semiclassical order, the near-horizon $\mathrm{Rindler}_2 \times S^2$ structure remains intact. Quantum corrections renormalize the surface gravity but do not alter its geometric origin. As long as the perturbative parameter $\eta \ll 1$, the horizon remains non-extremal and the semiclassical description is self-consistent. Thus semiclassical backreaction quantitatively modifies thermodynamic quantities while preserving the qualitative structure of Killing horizons.

The semiclassical correction to the heat capacity provides a direct measure of how quantum backreaction modifies the thermodynamic response of the system. To first order in $\eta$, the canonical heat capacity at fixed cavity radius retains the classical sign structure,
\[
C_{r_B}
=
C_{r_B}^{(0)}
\left[1 - \eta\,\Delta(x_B;a,k)\right],
\]
so that small black holes remain unstable and sufficiently large black holes remain stable within the canonical ensemble. The correction renormalizes the magnitude of the response, with explicit dependence on the cavity parameter $x_B$ and the stress--tensor coefficients. 

While the qualitative stability structure is unchanged at leading order, the location of the critical point receives a calculable $\mathcal{O}(\eta)$ shift. Semiclassical backreaction therefore preserves canonical stability in structure but modifies it quantitatively in a manner directly traceable to the integrated properties of the Hartle--Hawking stress tensor.

The regime of validity is controlled by the requirement $\eta \ll 1$, corresponding to black hole masses well above the Planck scale for typical quantum field energy densities. In this regime, fractional corrections to the horizon radius and temperature remain perturbatively small and higher-order quantum gravitational effects can be neglected. If the cavity radius becomes very large, infrared contributions enhance the backreaction, underscoring the importance of boundary conditions in defining equilibrium semiclassical configurations.

Beyond its specific analytic results, the present model provides a transparent laboratory for equilibrium semiclassical geometry. The minimal polynomial ansatz may be systematically refined to incorporate higher-order corrections or additional matter fields. The present analytic framework can be naturally extended to charged (Reissner--Nordstr\"om) or rotating (Kerr) black holes, where the interplay between semiclassical backreaction and additional horizon parameters may yield nontrivial modifications of equilibrium thermodynamics and stability properties.

In summary, this study provides a self-consistent analytic framework for understanding how vacuum polarization in the Hartle--Hawking state renormalizes black hole thermodynamics within a finite cavity. The structural decomposition of the temperature shift, the explicit horizon displacement formula, and the perturbative renormalization of canonical stability clarify the interplay between semiclassical backreaction and gravitational thermodynamics, strengthening the conceptual bridge between quantum field theory in curved spacetime and equilibrium black hole physics.
\begin{widetext}
\begin{table}[htbp!]
\begin{center}
\resizebox{\textwidth}{!}{%
\begin{tabular}{llllll}
\hline
Reference/Approach &
Quantum State &
RSET Treatment &
Backreaction Solution &
Finite Cavity &
Canonical Stability \\
\hline

Candelas (1980) \cite{Candelas:1980zt} &
Hartle--Hawking &
Numerical evaluation &
No full analytic integration &
No &
Not analyzed \\

Page (1982) \cite{Page:1982fm} &
Hartle--Hawking &
Polynomial fit to numerical RSET &
No explicit cavity backreaction &
No &
Not analyzed \\

Anderson--Hiscock--Samuel (1995) \cite{Anderson:1994hg} &
Hartle--Hawking &
Numerical RSET &
Numerical semiclassical geometry &
No &
Not analytic \\

York (1985, 1986) \cite{York:1984wp,York:1986it} &
Classical &
No quantum stress tensor &
Exact classical cavity solution &
Yes &
Canonical stability threshold $x_B = 2/3$ \\

Brown--York (1993) \cite{Brown:1992br} &
Classical &
No quantum stress tensor &
Quasilocal energy formalism &
Yes &
Classical thermodynamics \\

Present work &
Hartle--Hawking &
Minimal analytic model &
Closed-form analytic integration &
Yes &
Explicit $O(\eta)$ shift of stability threshold \\

\hline
\end{tabular}
}
\end{center}
\caption{Comparison between the present analytic semiclassical cavity model and representative treatments of semiclassical backreaction in Schwarzschild spacetime.}
\end{table}
\end{widetext}
\subsection*{Acknowledgments}
 This work was supported and funded by the Deanship of Scientific Research at Imam Mohammad Ibn Saud Islamic University (IMSIU)
(grant number IMSIU-DDRSP2602).

\end{document}